\documentstyle[aps,prd,graphicx]{revtex}

\begin{document}

%%%\draft
\title{Measurement of \mathversion{bold}$\psi(2S)$\mathversion{normal} 
decays to baryon pairs}
\author{
J.~Z.~Bai,$^1$      Y.~Ban,$^{11}$      J.~G.~Bian,$^1$
I.~Blum,$^{19}$
A.~D.~Chen,$^1$     G.~P.~Chen,$^1$     H.~F.~Chen,$^{18}$  H.~S.~Chen,$^{1}$
J.~Chen,$^5$
J.~C.~Chen,$^1$
% ...
\\(BES Collaboration)\\ }
\address{
$^1$ Institute of High Energy Physics, Beijing 100039, People's Republic of
     China\\
% ...
$^5$ Colorado State University, Fort Collins, Colorado 80523\\
% ...
$^{11}$ Peking University, Beijing 100871, People's Republic of China\\
% ...
$^{18}$ University of Science and Technology of China, Hefei 230026,
        People's Republic of China\\
$^{19}$ University of Texas at Dallas, Richardson, Texas 75083-0688\\
% ...
}

\date{\today}
\maketitle

\begin{abstract}
A sample of 3.95M $\psi(2S)$ decays registered in the BES detector are
used to study final states containing pairs of octet and decuplet baryons.
We report branching fractions for 
$\psi(2S)\to p\overline{p}$, $\Lambda\overline{\Lambda}$, 
$\Sigma^0\overline{\Sigma}{}^0$, $\Xi^-\overline{\Xi}{}^+$, 
$\Delta^{++}\overline{\Delta}{}^{--}$,
$\Sigma^+(1385)\overline{\Sigma}{}^-(1385)$, 
$\Xi^0(1530)\overline{\Xi}{}^0(1530)$, and $\Omega^-\overline{\Omega}{}^+$.
These results are compared to expectations based on 
the $SU(3)$-flavor symmetry, factorization, and perturbative QCD.

\end{abstract}

%%\narrowtext
% Begin main paper here
\section{Introduction}
In the quarkonium model, the $\psi(2S)$ is the first radial
excitation of the $^3S$ $c\overline{c}$ bound state.  As such, its
properties are expected to be relatively straight-forward to
understand, at least in terms of those of the $J/\psi$ ground state.
Somewhat surprisingly, these expectations do not always hold.  In
particular, there is a rather dramatic anomaly associated with
the $\psi(2S)$.

The major puzzle in hadronic $\psi$ decays is the large discrepancy
between the decay widths for $J/\psi(1S)\to\rho\pi$ and $K^*K$ and the
corresponding widths for $\psi(2S)$ decays.  These modes are expected
to proceed via $\psi\to ggg$, with widths that are proportional to the
square of the $c\overline{c}$ wave function at the origin, which is
well determined from dilepton decays.  The predicted ratio of branching
fractions from factorization is:
\begin{eqnarray*}
  \frac{{\cal B}(\psi(2S)\to X_{had})}{{\cal B}(J/\psi\to X_{had})}
   & = & \left[\frac{\alpha_s(\psi(2S))}{\alpha_s(J/\psi)}\right]^3
         \frac{{\cal B}(\psi(2S)\to e^+e^-)}{{\cal B}(J/\psi\to e^+e^-)} \\
   & = & 0.116 \pm 0.022,
\end{eqnarray*}
where $X_{had}$ designates any exclusive hadronic decay channel.  The
$\alpha_s^3$ terms come in from the three gluon
widths.~\cite{hadratio} Experimentally, the $\psi(2S)\to \rho\pi$ and
$K^*K$ are reduced by over a factor of twenty from these
expectations~\cite{rhopi}.  This anomaly calls into question the
underlying assumption behind the theoretical predictions: that the
$\psi(2S)$ is a pure $c\overline{c}$ state.

\subsection{$\psi(2S)\to B_i\overline{B_i}$}
\begin{figure}
  \centering
  \includegraphics[width=0.90\linewidth]{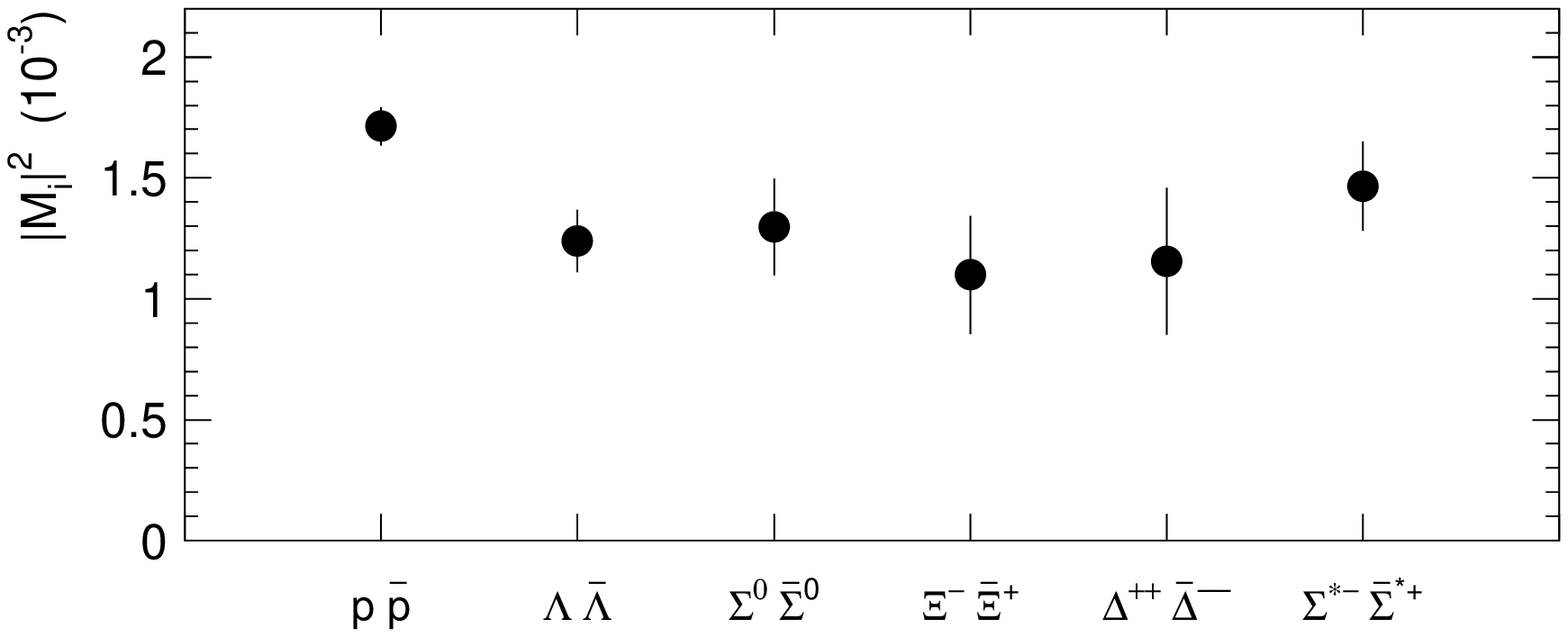}
  \caption{PDG values for the reduced branching fractions 
              $\left| M_i\right|^2 = 
              {\cal B}(J/\psi\to B_i\overline{B_i})
              /(\pi p^*/\sqrt{s})$ for $J/\psi\to
              B_i\overline{B_i},$ where $B_i \in \left\{ p, \Lambda,
              \Sigma^0, \Xi^-, \Delta^{++}, \Sigma^{\pm}(1385)\right\}$.}
  \label{fig:psi_rbf} 
\end{figure}

In the context of flavor $SU(3)$, a pure $c\overline{c}$ state is a
flavor singlet and, in the limit of $SU(3)$ flavor symmetry, the
phase-space-corrected reduced branching fractions to any baryon octet
pair, $\left| M_i \right|^2,$ where
$$
  \left| M_i \right| ^2 = \frac{{\cal B}(\psi(2S)\to
  B_i\overline{B_i})}{\pi p^*/\sqrt{s}}
$$
($p^*$ is the momentum of the baryon in the $\psi(2S)$ rest frame),
should be the same for every octet baryon, $B_i$.  Deviations from
this rule could indicate a non-$c\overline{c}$ component of the
charmonium wave function.  The reduced branching fractions for
$J/\psi\to B_i\overline{B_i}$ decays are shown in
Fig.~\ref{fig:psi_rbf}.  The $SU(3)$ relation works reasonably well,
although there may be some increase for the $p\overline{p}$ mode.

This relation has not been tested for the $\psi(2S)$, where the
only relevant mode that has been measured is $p\overline{p}$, and that
with rather poor precision~\cite{dasp},\cite{marki}.

There are very few direct calculations of the decay of charmonium to
baryonic final states.  One of the most comprehensive is the
perturbative analysis by Bolz and Kroll~\cite{bk}.  A comparison to
this analysis will be discussed later.

\section{This Experiment}

We report results of measurements
of the branching fractions for $\psi(2S)\to B_i\overline{B_i},$
where $B_i \in \{ p, \Lambda, \Sigma^0, \Xi^-, $ $\Delta^{++},
\Sigma^{+}(1385), \Xi^0(1530), \Omega^-\}$ using a sample of
$3.95\times 10^6$ $\psi(2S)$ events produced via $e^+e^-$
annihilations at the BEPC collider and observed by the BEijing Spectrometer (BES).
The data represents a total integrated luminosity of
$\approx 6.7$~\mbox{pb${}^{-1}$}.

The Beijing Electron Spectrometer, BES, is a conventional cylindrical
magnetic spectrometer, coaxial with the BEPC colliding $e^+e^-$
beams~\cite{detector}.  A four-layer central drift chamber (CDC)
surrounding the beampipe provides trigger information.  Outside the
CDC, a forty-layer main drift chamber (MDC) provides tracking and
energy-loss ($dE/dx$) information on charged tracks over $85\%$ of the
total solid angle.  The momentum resolution is $\sigma _p/p = 0.017
\sqrt{1+p^2}$ ($p$ in \mbox{GeV$/c$}), and the $dE/dx$ resolution for
hadron tracks is $\approx 11\%$.  An array of 48 scintillation
counters surrounding the MDC provides time-of-flight (TOF) information
of charged tracks with a resolution of $\approx 450$~\mbox{ps} for
hadrons. Outside the TOF system, a 12 radiation length, lead-gas
barrel shower counter (BSC), operating in self-quenching streamer
mode, measures the energies of electrons and photons over $\approx
80\%$ of the total solid angle. The energy resolution is
$\sigma_E/E=0.22/\sqrt{E}$ ($E$ in \mbox{GeV}), and the spatial
resolutions are $\sigma _\phi = 4.5$~mrad and
$\sigma_z=4$~\mbox{cm}. Surrounding the BSC is a solenoidal magnet
that provides a 0.4~Tesla magnetic field in the central tracking
region of the detector. Three double layers of planar counters
instrument the magnet flux return (MUID) and are used to identify
muons of momentum greater than 0.5 \mbox{GeV$/c$}. Endcap
time-of-flight and shower counters extend coverage to the forward and
backward regions.

\section{Baryon Octet}
\subsection{$\boldmath\psi(2S)\to p\overline{p}$}
\label{sec:psip_pp}

The experimental signature for the decay $\psi(2S)\to p\overline{p}$
is two back-to-back, oppositely charged tracks each with a momentum of
1.586~\mbox{GeV$/c$}.  The proton typically deposits one-half or less
of its 0.91~\mbox{GeV} kinetic energy in the BSC; the antiproton
undergoes an annihilation process in the BSC approximately half the
time, producing a large shower.

Major potential  backgrounds are:  $\psi(2S)\to K^+K^-$,
$\pi^+\pi^-$, $\mu^+\mu^-$, and $e^+e^-$.  Each of these modes has a
momentum at least 190~\mbox{MeV$/c$} greater than that of the
$p\overline{p}$ channel.  

We select events with two
and only two well reconstructed, oppositely charged tracks with good
time of flight information, and which are not identified as muons by
the muon system.  Also $\left|\cos\theta\right|$ must be less than 0.6
for both tracks to ensure that they occur within the fiducial volume
covered by the muon system.  Candidate $p\overline{p}$ pairs are
required to be within 1.8 degrees of collinear.

\begin{figure}
  \centering
  \includegraphics[width=0.90\linewidth]{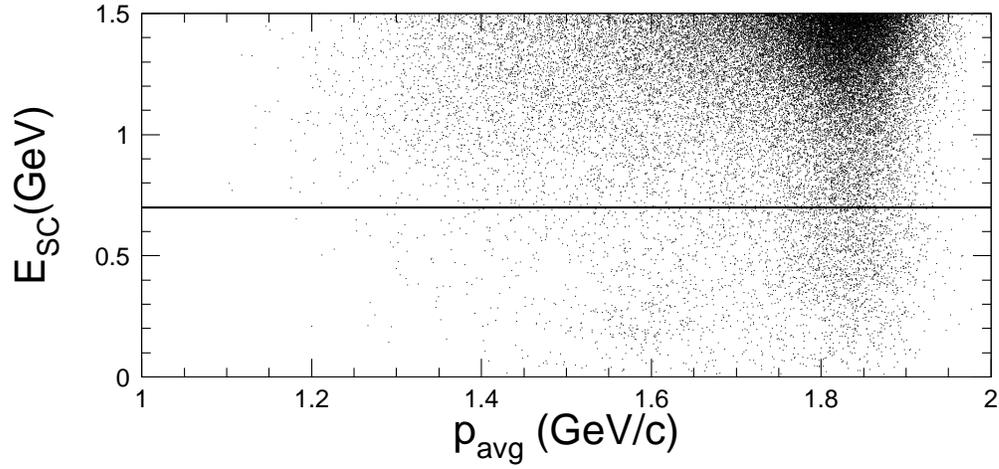}
  \caption{Shower counter energy \textit{vs.} momentum for positively 
              charged tracks.  Signal is expected near a momentum of
              1.6~\mbox{GeV/$c$}, and energy of 300~\mbox{MeV}.}
  \label{fig:pp_esc} 
\end{figure}

The shower counter energy deposition as a function of momentum for
positively charged tracks is shown in Figure~\ref{fig:pp_esc}.  The
faint cluster near $p=1.6~\mbox{GeV$/c$}$, $E_{SC}=0.3~\mbox{GeV}$ is
the proton signal.  The other features on the graph are due to
Bhabhas (large concentration at $p\approx1.8~\mbox{GeV$/c$}$,
$E_{SC}\approx 1.5~\mbox{GeV}$), muons (vertical stripe at
$p\approx1.8~\mbox{GeV$/c$}$, $E_{SC} < 1~\mbox{GeV}$) and radiative
Bhabhas (trailing cluster at $p<1.6~\mbox{GeV$/c$}$, $E_{SC}\approx
p$).  To remove these backgrounds, a cut is made at
$E_{SC}<0.7~\mbox{GeV$/c$}$.  In addition, the shower counter has a
number of support ribs which are dead regions, thus degrading the
energy measurement.  Tracks which enter these regions are removed from
consideration.

\begin{figure}
  \centering
  \includegraphics[width=0.90\linewidth]{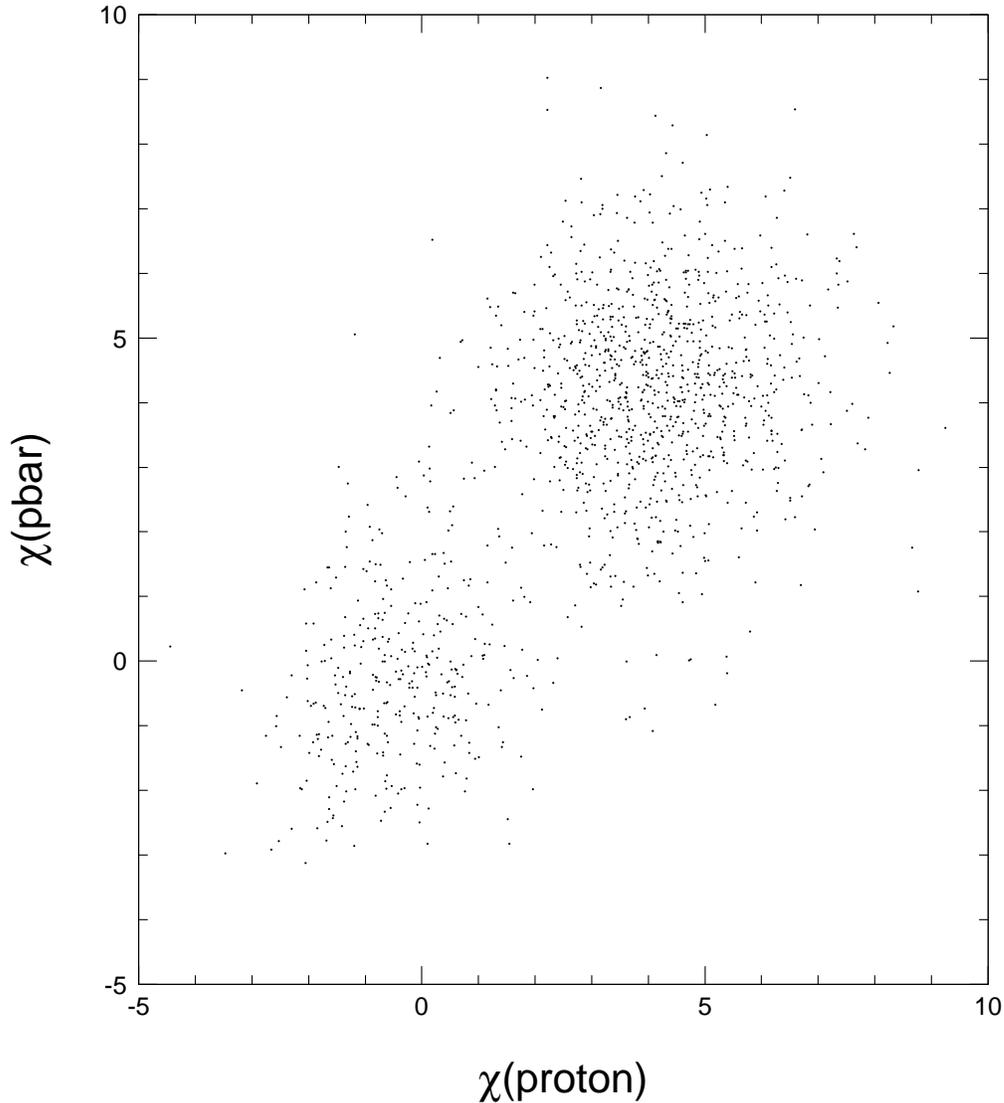}
  \caption{Distribution of $\chi_p$ for antiproton candidates 
              versus $\chi_{\overline{p}}$ for proton candidates, where
              $\chi_p =\left| dE/dx_{\rm meas} - \right.$ 
              $\left. dE/dx_{\rm exp}\right|$
              and is calculated assuming the track to be a proton (or
              anti proton).  The signal is the cluster near (0,0).  The
              cut made is $(\chi^2_p + \chi^2_{\overline{p}})^\frac{1}{2}
              < 3$.}
  \label{fig:pp_xsp}
\end{figure}

An additional handle on the identification of protons is gained from
the $dE/dx$ system.  Figure~\ref{fig:pp_xsp} shows the $dE/dx$
particle ID results for candidate events that pass the above cuts.
Units are $\chi = \left| dE/dx_{\rm meas} - dE/dx_{\rm exp}\right|
/\sigma$, where $\sigma$ is the resolution of the particle ID system.
The vertical axis is for the $\overline{p}$ hypothesis, and the
horizontal refers to the $p$ hypothesis.  The cluster near (0,0)
contains true $p\overline{p}$ events, and the cluster near (5,5) is a
mixture of event types such as radiative Bhabhas and $\psi^\prime\to
ee$.  A cut is made on the combined $\chi$, $(\chi^2_p
+\chi^2_{\overline{p}})^\frac{1}{2} < 3$.

\begin{figure}
  \centering
  \includegraphics[width=0.90\linewidth]{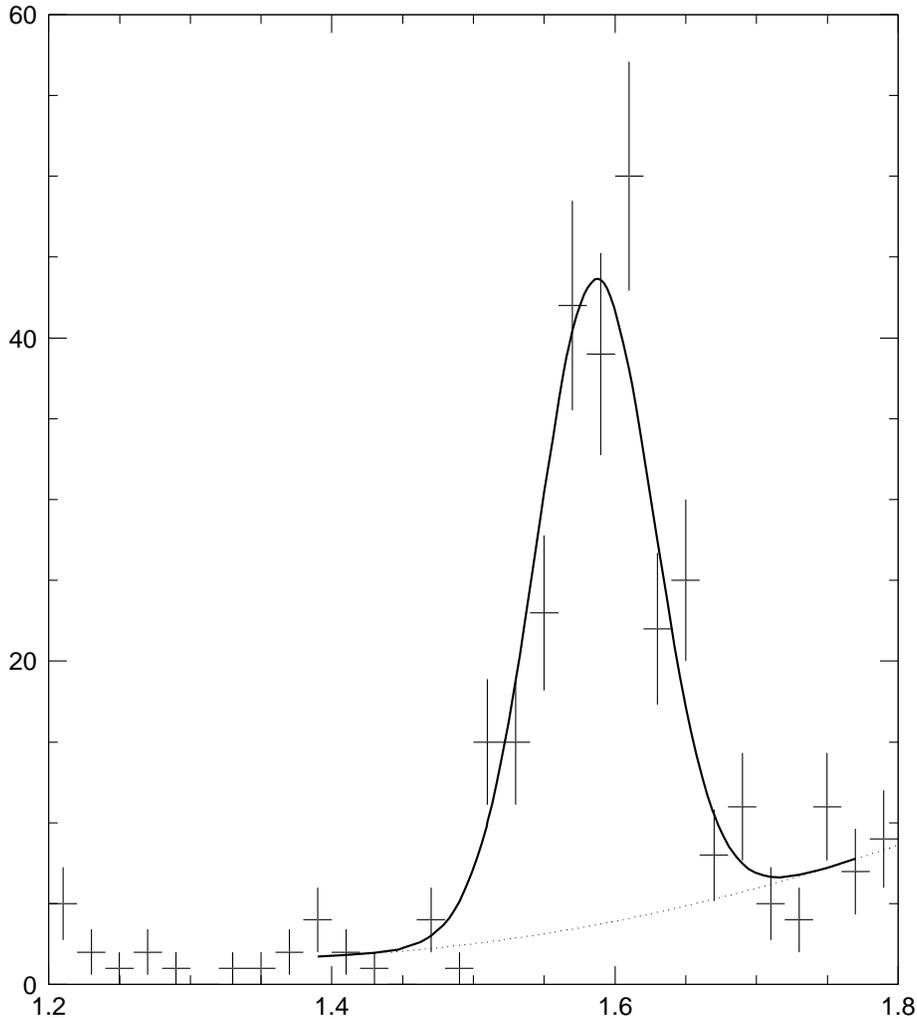}
  \caption{Weighted average momentum of $p\overline{p}$ pairs, 
              fit to a gaussian plus a quadratic.}
\label{fig:pp_fit} 
\end{figure}

The weighted average momentum spectrum of the remaining candidate
events is shown in Figure~\ref{fig:pp_fit}.  By weighted average we
mean that the track parameters of the positive and negative tracks
(curvature and dip-angle) are averaged together and then combined to
form a momentum.  This spectrum in Figure~\ref{fig:pp_fit} is fit to a
gaussian plus a quadratic background function, with the centroid of
the gaussian fixed to the theoretic momentum of the protons,
1.586~\mbox{GeV$/c$}.  The width and height are allowed to vary.  From
the fit, $N_{\mathstrut p\overline{p}}= 201 \pm 14 \pm 20$.  Here and
below, the first error is statistical and the second is systematic, in
this case the error on the fit.

\subsection{$\psi(2S)\to \Lambda\overline{\Lambda}$}
\label{sec:psip_ll}

The decays $\psi(2S)\to\Lambda\overline{\Lambda}$ produce two
back-to-back $\Lambda$s, each with momentum $1.467$~\mbox{GeV$/c$}.  We
only consider events where both $\Lambda$s decay to the charged $p\pi$
final state.  The final states of interest are thus, $\psi(2S)\to
\pi^+\pi^- p\overline{p}$, where the $p\pi^-$ and $\overline{p}\pi^+$
originate from well separated decay vertices.  The decay kinematics are
such that the proton (antiproton) is always the highest momentum
positive (negative) track in the event.

\begin{figure}
  \centering
  \includegraphics[width=0.90\linewidth]{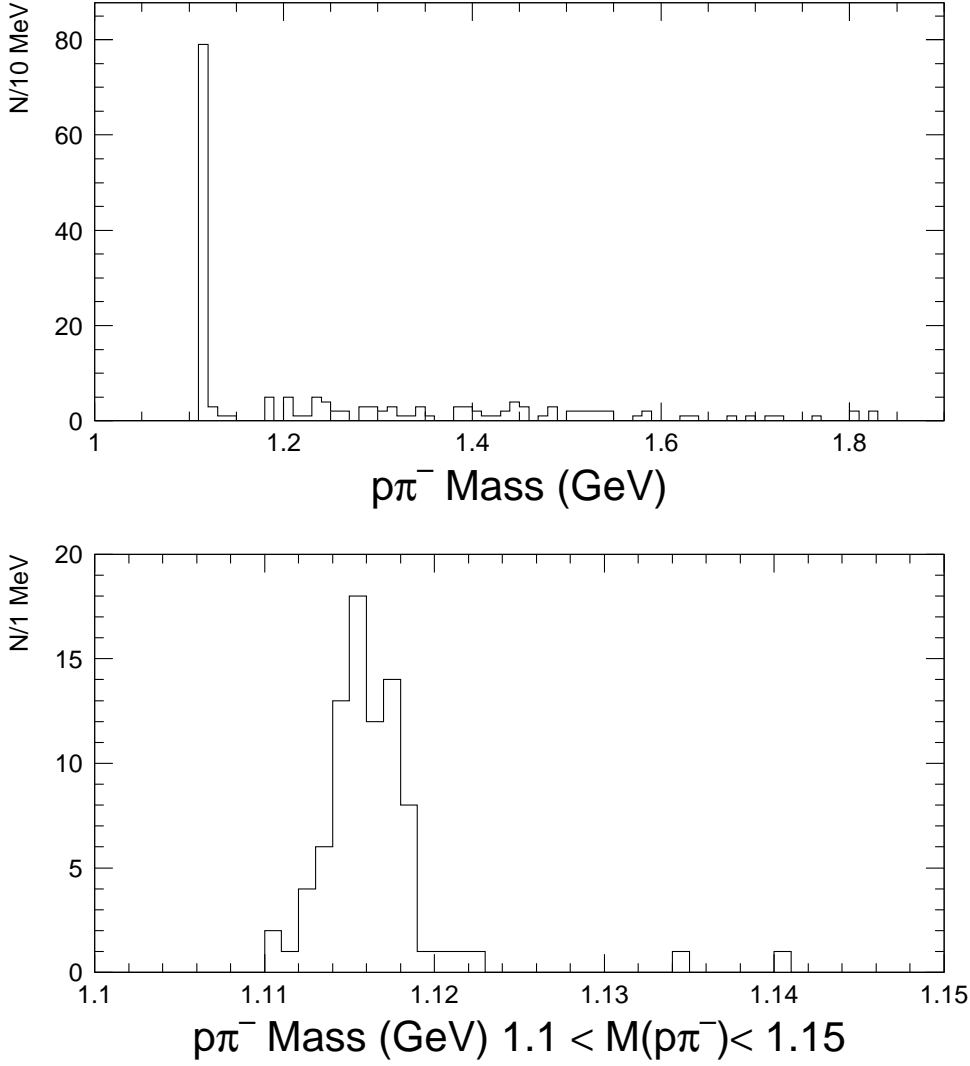}
  \caption{Distribution of $p\pi$ invariant masses from a kinematic fit to
              $\psi(2S)\to p\overline{p}\pi^+\pi^-$, $M_{p\pi^-} =
              M_{\overline{p}\pi^+}$.  Upper figure is full range of
              $M_{p\pi^-}$, lower figure is expanded near the signal peak
              at 1.11~$\mbox{GeV}/c^2$.}
  \label{fig:llmass} 
\end{figure}

We select events with four and only four well reconstructed tracks
with a zero net charge, and in the fiducial region covered by the
drift chamber, $\left|\cos\theta\right|\le 0.80$.  Events which pass
these cuts are processed through a detached vertex finding algorithm,
and subjected to a 5-C kinematic fit to $p\overline{p}\pi^+\pi^-$,
with $M_{p\pi^-}=M_{\overline{p}\pi^+}$.  The 84 events which pass
this fit with a confidence level of more than 1\%, 
and have $M_{p\pi}
\le 1.15$~\mbox{GeV$/c^2$} are shown in Figure~\ref{fig:llmass}.
Extrapolating the two events in the region above 1.13~\mbox{GeV$/c^2$}
and below 1.15~\mbox{GeV$/c^2$} to the area under the mass peak, we
find that there are four background events in the plot.  We
conservatively assign this number a 100\%{} error and determine
$N_{\mathstrut\Lambda\overline{\Lambda}}$ to be $80 \pm 9 \pm 4$.

\subsection{$\psi(2S)\to \Sigma^0\overline{\Sigma}{}^0$}
\label{sec:psip_soso}

The $\Sigma^0$ hyperons from $\psi(2S)\to
\Sigma^0\overline{\Sigma}{}^0$ decay promptly via
$\Sigma^0\to\gamma\Lambda$. We consider only those decays where the
daughter $\Lambda$s decay via the charged $p\pi$ mode. The
experimental signature is thus, $\psi(2S)\to
p\overline{p}\pi^+\pi^-\gamma\gamma$, where the $p\pi^-$ and
$\overline{p}\pi^+$ originate from $\Lambda$ hyperons with well
separated decay vertices.  In addition, there are two photons in the
energy range $27\le E_{\gamma}\le 202$~\mbox{MeV}.  As in the case for
$\psi(2S)\to\Lambda\overline{\Lambda},$ the proton (antiproton) is
always the highest momentum positive (negative) track in the final
state.

We extract $\psi(2S)\to \Sigma^0\overline{\Sigma}{}^0$ event candidates 
($\Sigma^0\to\Lambda\gamma$, $\Lambda\to p\pi^-$) using the same selection 
criteria as used for the $\Lambda\overline{\Lambda}$ mode with the 
additional requirement that there be two or more isolated clusters in the 
BSC with energy greater than 60~\mbox{MeV}, and within region 
$\left|\cos\theta\right|\le 0.75$.  By \textit{``isolated''} we mean more 
than $12.8^\circ$ ($\cos\theta_{\rm isol}<.975$) away from each of the 
charged tracks.

Both $p\pi$ pairs in the surviving events are processed through a displaced 
vertex-finding algorithm and the event is then subjected to a 
five-constraint kinematic to the hypothesis $\psi(2S)\to 
p\overline{p}\pi^+\pi^-\gamma\gamma$, with the beam constraint 
$M_{p\pi^-\gamma} = M_{\overline{p}\pi^+\gamma}$.  Here the highest 
momentum positive (negative) track is classified as the proton 
(antiproton).  For events with more than two $\gamma$ candidates, the fit 
is applied for each possible combination.

\begin{figure}
  \centering
  \includegraphics[width=0.90\linewidth]{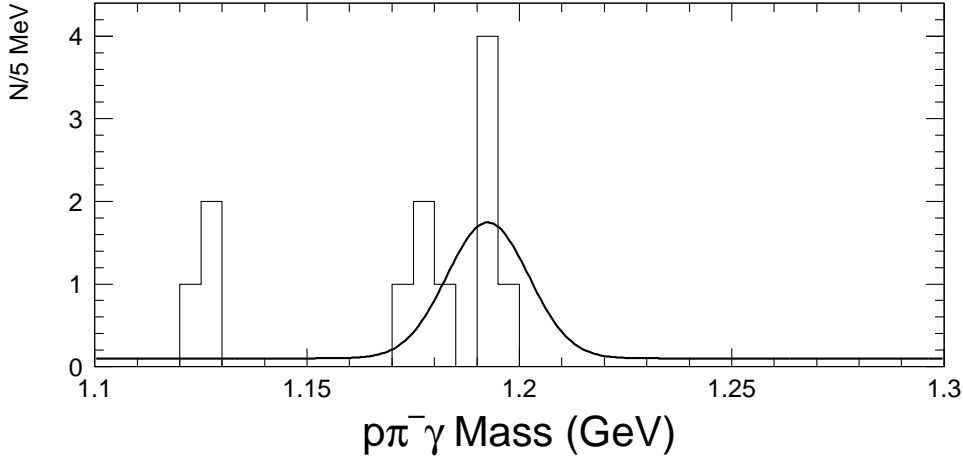}
  \caption{\label{fig:sigfit} 
              Distribution of $p\pi\gamma$ invariant masses from a
              kinematic fit to $\psi(2S)\to
              p\overline{p}\pi^+\pi^-\gamma\gamma$, $M_{p\pi^-\gamma} =
              M_{\overline{p}\pi^+\gamma}$.  Events with $p\pi\gamma$
              masses below 1.3~$\mbox{GeV}/c^2$ are fit to a gaussian
              signal plus a linear background.  There are $8 \pm 3 \pm 2$
              events in the $\Sigma^0$ peak.}
\end{figure}

Events which pass the kinematic fit with a confidence level greater
than 1\%, 
and $M_{\mathstrut p\pi\gamma}<1.3~\mbox{GeV}/c^2$ are shown
in Figure~\ref{fig:sigfit}.  We fit this spectrum to a single
gaussian plus a linear background with the peak fixed to the mass of
the $\Sigma^0$, 1.192~\mbox{GeV}$/c^2$.  From the fit,
$N_{\mathstrut\Sigma^0\overline{\Sigma}{}^0} = 8 \pm 3
\pm 2$.

\subsection{$\psi(2S)\to \Xi^-\overline{\Xi}{}^+$}
\label{sec:psip_xx}

The $\Xi^-$ hyperon from $\psi(2S)\to \Xi^-\overline{\Xi}{}^+$ decays 
via $\Xi^-\to\pi^-\Lambda$.  We consider only those decays where the 
daughter $\Lambda$s decay via the charged $p\pi$ mode.  The experimental 
signature is thus $\psi(2S)\to p\overline{p}\pi^+\pi^-\pi^+\pi^-$ where 
one each of the $p\pi^-$ and $\overline{p}\pi^+$ combinations originate 
from $\Lambda$ hyperons with well separated decay vertices.  As in the case 
for $\psi(2S)\to\Lambda\overline{\Lambda}$ and 
$\Sigma^0\overline{\Sigma}{}^0$, the proton (antiproton) is always the 
highest momentum positive (negative) track in the final state.

We select events with six and only six well reconstructed tracks with
zero net charge, and in the fiducial region covered by the drift
chamber, $\left|\cos\theta\right|\le 0.80$.  
Each of the four possible $p\pi^-$ and $\overline{p}\pi^+$ combinations are 
sent through a displaced vertex-finding algorithm and subsequently 
subjected to a five-constraint kinematic fit to the hypothesis 
$\psi(2S)\to p\overline{p}\pi^+\pi^-\pi^+\pi^-$, with the beam 
constraint $M_{p\pi^-\pi^-} = M_{\overline{p}\pi^+\pi^+}$.  

Events which pass the fit with a confidence level greater than 1\%
are examined further.  We additionally require that the 
$p\pi$ combinations have a mass within 10~\mbox{MeV}$/c^2$ of
the $\Lambda$ and that the mass of the $\Lambda\overline{\Lambda}$
candidate is more than 20~\mbox{MeV}$/c^2$ away from the $J/\psi$ in order
to reduce background from the cascade decay $\psi(2S)\to J/\psi\pi\pi$,
$J/\psi\to\Lambda\overline{\Lambda}$.  

\begin{figure}
  \centering
  \includegraphics[width=0.90\linewidth]{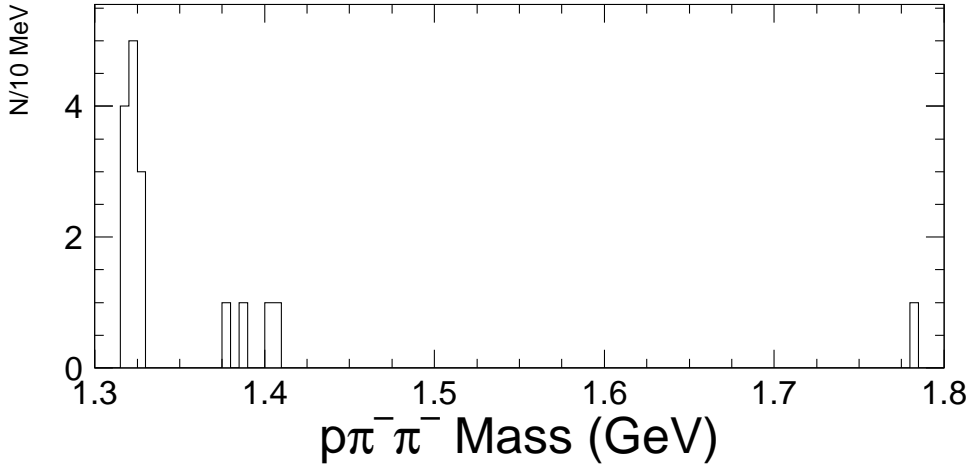}
  \caption{Distribution of $p\pi^-\pi^-$ masses from a kinematic fit to 
              $\psi(2S)\to p\overline{p}\pi^+\pi^-\pi^+\pi^-$,
              $M_{p\pi^-\pi^-}=M_{\overline{p}\pi^+\pi^+}$.
              There are $12 \pm 3.4\pm 0.2$ events in the $\Xi^-$ peak.}
  \label{fig:ximass} 
\end{figure}

The $M_{\mathstrut p\pi^-\pi^-}$ spectrum of events which remain after 
the above cuts is plotted 
in Figure~\ref{fig:ximass}.  There are $12\pm 3.4$ events in the $\Xi^-$ 
peak.  Averaging the five events outside the peak region over the entire 
plot and multiplying by the width of the signal gives 0.15 background 
events.  A conservative error of 100 percent is applied, giving $12 \pm 
3.4 \pm 0.2$ $\psi(2S)\to\Xi^-\overline{\Xi}{}^+$ events detected.

\section{Baryon Decuplet}
\subsection{$\psi(2S)\to\Delta^{++}\overline{\Delta}{}^{--}$}
\label{sec:psip_dd}

The decay $\psi(2S)\to\Delta^{++}\overline{\Delta}{}^{--}$ 
produces back-to-back $\Delta^{++}$ and $\overline{\Delta}{}^{--}$.
As the $\Delta^{++}$ is a broad (111~\mbox{MeV}$/c^2$) resonance,
the primary hyperons do not have well-defined momenta, in contrast
to the octet cases above.
We select events where both $\Delta^{++}$ and
$\overline{\Delta}{}^{--}$ decay to $p\pi$ [${\cal B}(\Delta^{++}\to
p\pi) > 99 \%$].  The final state is $\psi(2S)\to
p\overline{p}\pi^+\pi^-$.

We select events with four and only four well reconstructed tracks with a 
zero net charge, and in the fiducial region covered by the drift chamber, 
$\left|\cos\theta\right|\le 0.80$.  The surviving events are processed 
through a four-constraint kinematic fit to the hypothesis $\psi(2S)\to 
p\overline{p}\pi^+\pi^-$.  Events which pass with a confidence level 
greater than 1\%\ are examined further.  

\begin{figure}
  \includegraphics[width=0.90\linewidth]{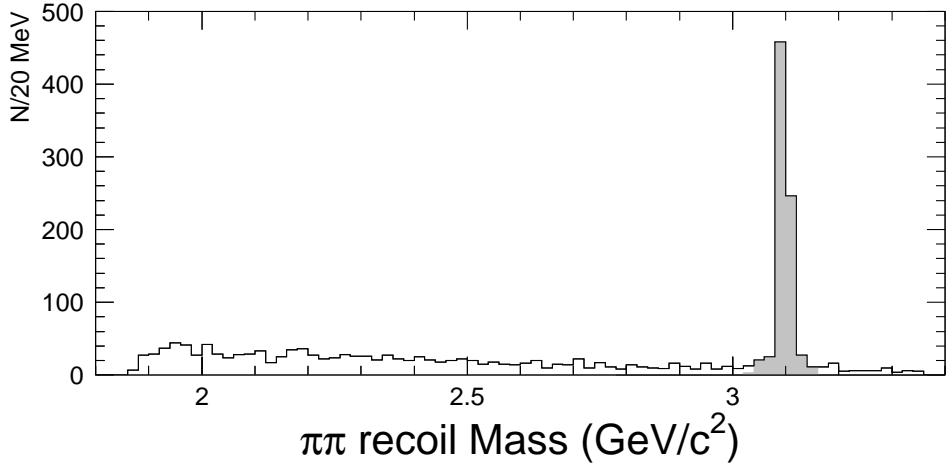}
  \caption{Distribution of $\pi^+\pi^-$ recoil masses in 
              the $\Delta^{++}$ analysis.  A cut is made at
              $\left|M_{p\overline{p}}-M_{J/\psi}\right|>
              60~\mbox{MeV}/c^2$
              to remove $J/\psi$ contamination.}
  \label{fig:dpp_psi}
\end{figure}

\begin{figure}
  \includegraphics[width=0.90\linewidth]{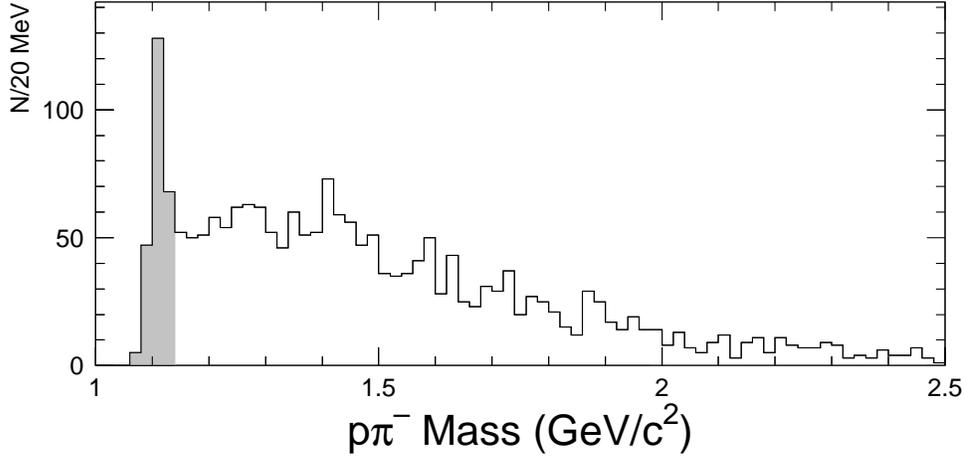}
  \caption{Distribution of $p\pi^-$ masses in the $\Delta^{++}$ analysis.  
              A cut is made at $M_{ p\pi^-}>1.14 $~\mbox{GeV}$/c^2$ to
              remove $\Lambda\overline{\Lambda}$ background.}
  \label{fig:dpp_lam}
\end{figure}

Figure~\ref{fig:dpp_psi} shows the invariant mass distribution of the 
$p\overline{p}$ pair in events which pass the fit.  There is a clear peak 
in the $J/\psi$ mass region coming from the cascade decay $\psi(2S)\to 
J/\psi\pi^+\pi^-$, $J/\psi\to p\overline{p}$; we remove this by making a 
60~\mbox{MeV}$/c^2$ cut around the $J/\psi$.  Figure~\ref{fig:dpp_lam} 
shows the invariant mass distribution for $p\pi^-$ containing a peak at the 
$\Lambda$ mass.  We remove the $\Lambda\overline{\Lambda}$ background by 
requiring the $p\pi^-$ and $\overline{p}\pi^+$ masses to be greater than 
1.15~\mbox{GeV}$/c^2$.

\begin{figure}
  \centering
  \includegraphics[width=0.90\linewidth]{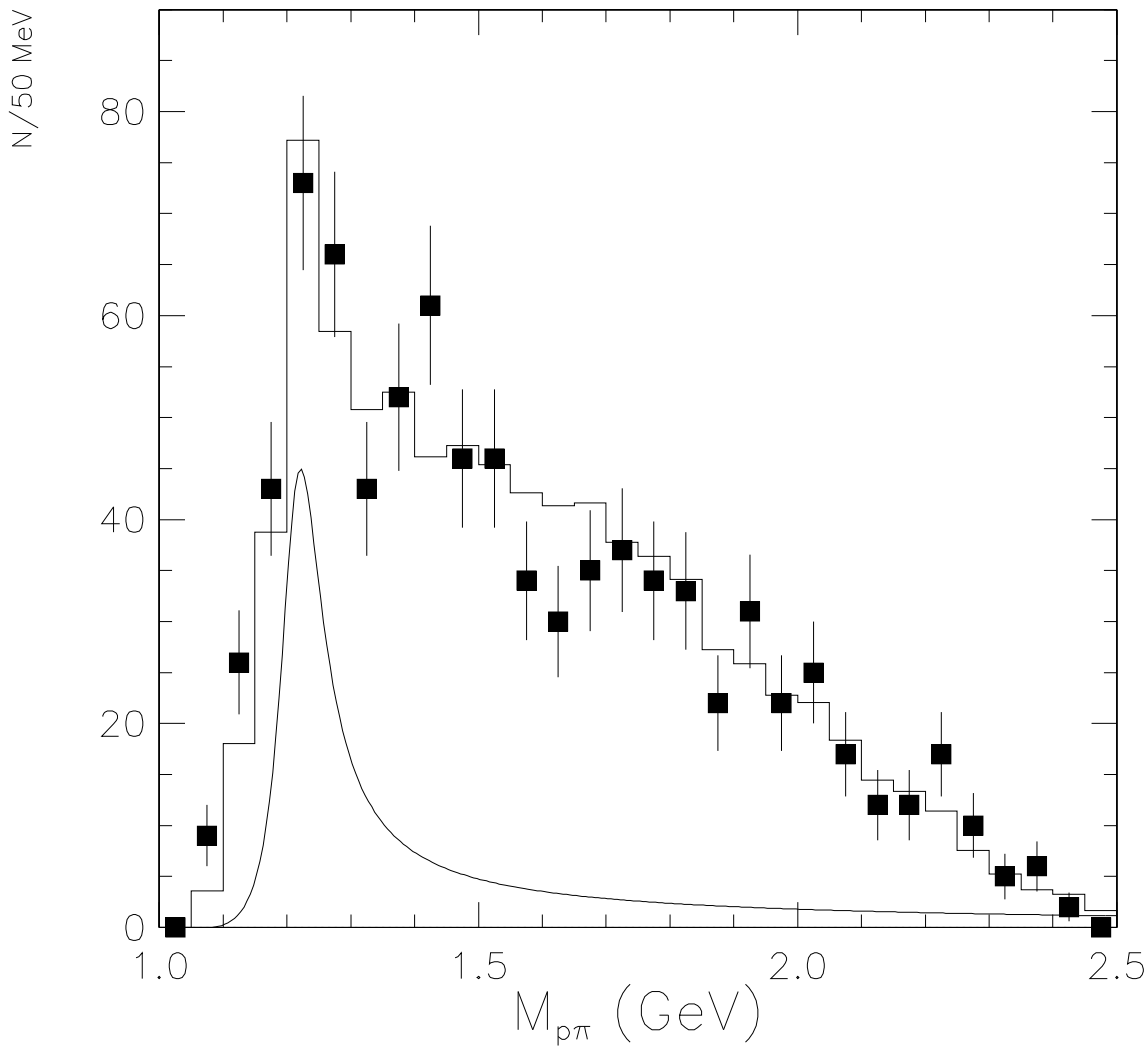}
  \caption{Distribution of  $p\pi^+$ masses from a kinematic fit to 
              $\psi(2S)\to p\overline{p}\pi^+\pi^-$.  Data is fit to a
              spin-1 Breit-Wigner plus a 4-body phase-space background
              (from Monte Carlo).  Black boxes with error bars are data,
              smooth curve is the spin-1 Breit-Wigner fit result, and
              histogram is the final fit to background plus Breit-Wigner,
              binned to match the data.
              $N_{\Delta^{++}\overline{\Delta}{}^{--}} = 157 \pm 13 \pm
              34$.}
  \label{fig:dppfit}
\end{figure}

Events which pass all above cuts are fit to a spin-1 Breit-Wigner plus
a 4-body phase-space background histogram.  The width and centroid of
the signal spectrum are fixed to the PDG\cite{PDG} values.
Figure~\ref{fig:dppfit} shows the output of the fit; there are 849
total events in the plot.  The fit parameter varied is the relative
proportions of the phase space background and the Breit-Wigner signal
to the total number of events in the plot.
$N_{\mathstrut\Delta^{++}\overline{\Delta}{}^{--}}$ is $157 \pm 13 \pm
34$.

\subsection{$\psi(2S) \to \Sigma^+(1385)\overline{\Sigma}{}^-(1385)$}
\label{sec:psip_stst}

The hyperons from $\psi(2S)\to \Sigma^{*+}\overline{\Sigma}{}^{*-}$
decay via $\Sigma^{+}(1385)\to\Lambda\pi^{+}$ 88\% of the time.  We
consider only those decays where the daughter $\Lambda$s decay via the
charged $p\pi$ mode. The experimental signature is $\psi(2S)\to
p\overline{p} 2(\pi^+\pi^-)$ where one each of the $p\pi^-$
($\overline{p}\pi^+$) candidates is consistent with being from the
decay of a $\Lambda$ ($\overline{\Lambda}$).

We select events with six and only six well reconstructed tracks with
a zero net charge, and in the fiducial region covered by the drift
chamber, $\left|\cos\theta\right|\le 0.80$.  We kinematically fit the
36 possible charge combinations of $(+-)+(-+)-$, running the
$(+-)/(-+)$ candidates through a displaced vertex finding algorithm,
to $p\pi^-\pi^+\overline{p}\pi^+\pi^-$.  No constraints are placed on
the $(+-)/(-+)$ candidates.

\begin{figure}
  \centering
  \includegraphics[width=0.90\linewidth]{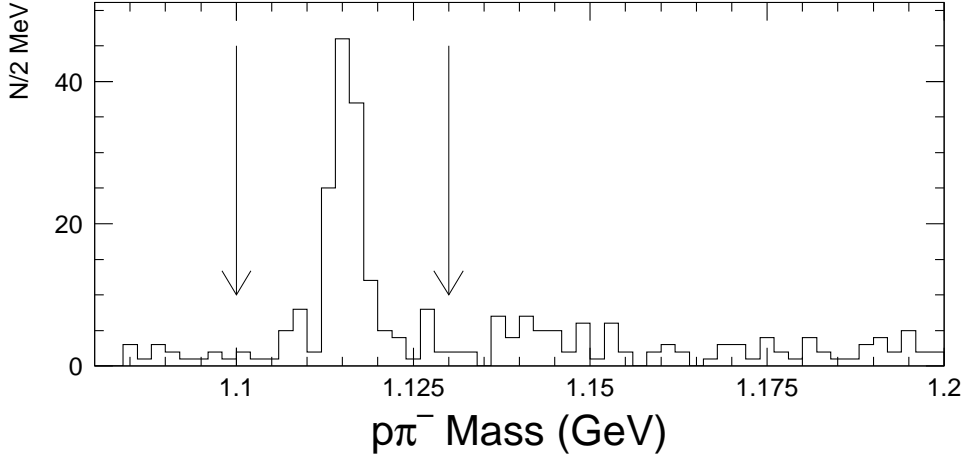}
  \caption{Distribution of $M_{p\pi^-}$ from a fit to $\psi(2S)\to$
              $(p\pi^-)\pi^+(\overline{p}\pi^+)\pi^-$, showing the $\Lambda$
              peak in $\Sigma(1385)^{+}\overline{\Sigma}(1385){}^{-}$
              candidate events.}
  \label{fig:sslam}     
\end{figure}

Figure~\ref{fig:sslam} shows that the fit mass of the daughter
$\Lambda$s from the decay of the primary $\Sigma^{*+}$ is well defined
and centered at the $\Lambda$ mass.  We make a loose cut of
15~\mbox{MeV}$/c^2$ on the $\Lambda$ and $\overline{\Lambda}$
resonances, indicated by the arrows on the plot.

\begin{figure}
  \centering
  \includegraphics[width=0.90\linewidth]{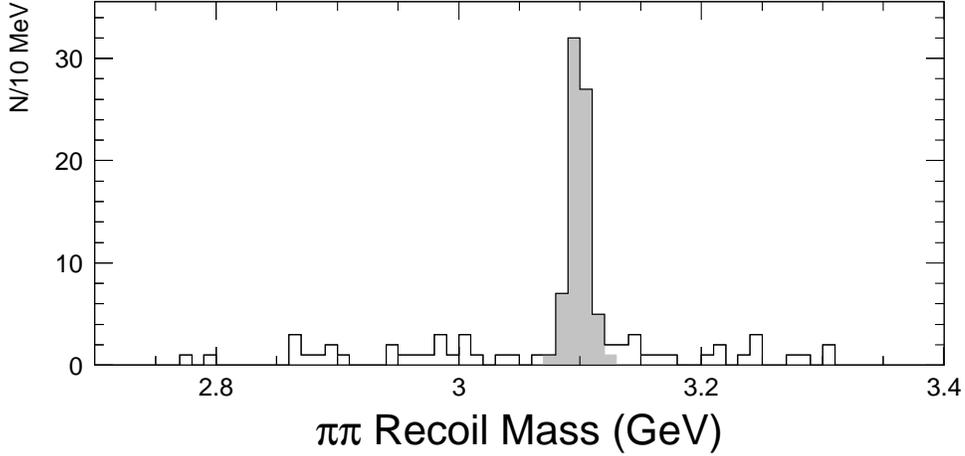}
  \caption{Distribution of $\pi^+\pi^-$ recoil masses from a kinematic fit
              to $\psi(2S)\to\Lambda\overline{\Lambda}\pi^+\pi^-$, showing
              large $J/\psi$ contamination in
              $\Sigma(1385)^{+}\overline{\Sigma}(1385){}^{-}$ candidate
              events.}
  \label{fig:sstar_pp}
\end{figure}

As shown in Figure~\ref{fig:sstar_pp}, the mass recoiling against the orphan
$\pi^+\pi^-$ pair is dominated by a peak at the $J/\psi$ mass, indicating
contamination of $\psi(2S) \to \pi^+\pi^-J/\psi$, $J/\psi\to 
\Lambda\overline{\Lambda}$.  We therefore remove events with a
$\pi^+\pi^-$ recoil mass within 30~\mbox{MeV}$/c^2$ of the $J/\psi$.

\begin{figure}
  \centering
  \includegraphics[width=0.90\linewidth]{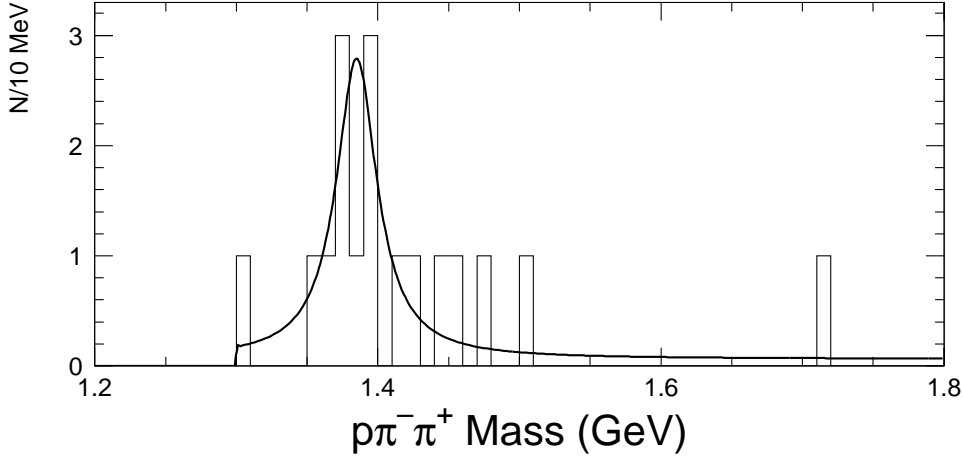}
  \caption{Distribution of $M_{p\pi^-\pi^-}$ from a kinematic fit to
              $\psi(2S)\to (p\pi^-)\pi^-(\overline{p}\pi^+)\pi^+$.
              The histogram is an unbinned fit to a Breit-Wigner
              constrained to the nominal $\Sigma(1385)^{+}$ mass and
              width.}
  \label{fig:sstar_mass} 
\end{figure}

To determine $N_{\mathstrut\Sigma^{*+}\overline{\Sigma}{}^{*-}}$, we 
constrain the $\overline{p}\pi^+\pi^-$ combination 
($\overline{\Sigma}{}^{*-}$ candidate) to be within 107.4~\mbox{MeV}$/c^2$ 
($3\times\Gamma$) of the nominal PDG value in order to enhance the 
$\Sigma^{*+}$ signal.  Events which pass the above cuts are fit to a 
Breit-Wigner with a constant background, with the mass and width fixed to 
the PDG values ($M=1382.8~\mbox{GeV}/c^2$, $\Gamma=35.8~\mbox{MeV}/c^2$).  
This fit is shown in Figure~\ref{fig:sstar_mass}; from the fit, 
$N_{\mathstrut\Sigma^{*+}\overline{\Sigma}{}^{*-}} = 13.8 \pm 3.7 \pm 2.7$.

\subsection{$\psi(2S) \to \Xi^0(1530)\overline{\Xi}{}^0(1530)$}
\label{sec:psip_xsxs}

In the decay $\psi(2S)\to\Xi^*\overline{\Xi}{}^*$,
the {$\Xi^*$}s are produced back to back in 
the $\psi(2S)$ rest frame.  The dominant decay mode of $\Xi^*$ baryons is 
$\Xi^*\to\Xi^-\pi^+$, with a branching fraction of 0.66.\cite{PDG} The
$\Xi^-$ decays as in Section~\ref{sec:psip_xx} to $\Lambda\pi^-$, and the
$\Lambda$ decays to $p\pi^-$.

We select events with eight and only eight well reconstructed tracks with 
zero net charge, and in the fiducial region covered by the drift chamber, 
$\left|\cos\theta\right|\le 0.80$.  Remaining events are subjected to a 
4-constraint kinematic fit to the hypothesis $\psi(2S) \to p\pi^- 
\overline{p}\pi^+ \pi^+\pi^-\pi^+\pi^-$.  The $p\pi$ candidates for $\Lambda$s 
are sent through a displaced vertex finder.  Events which pass the fit
with a fit probability greater than 0.01 are examined further.

\begin{figure}
  \centering
  \includegraphics[width=0.90\linewidth]{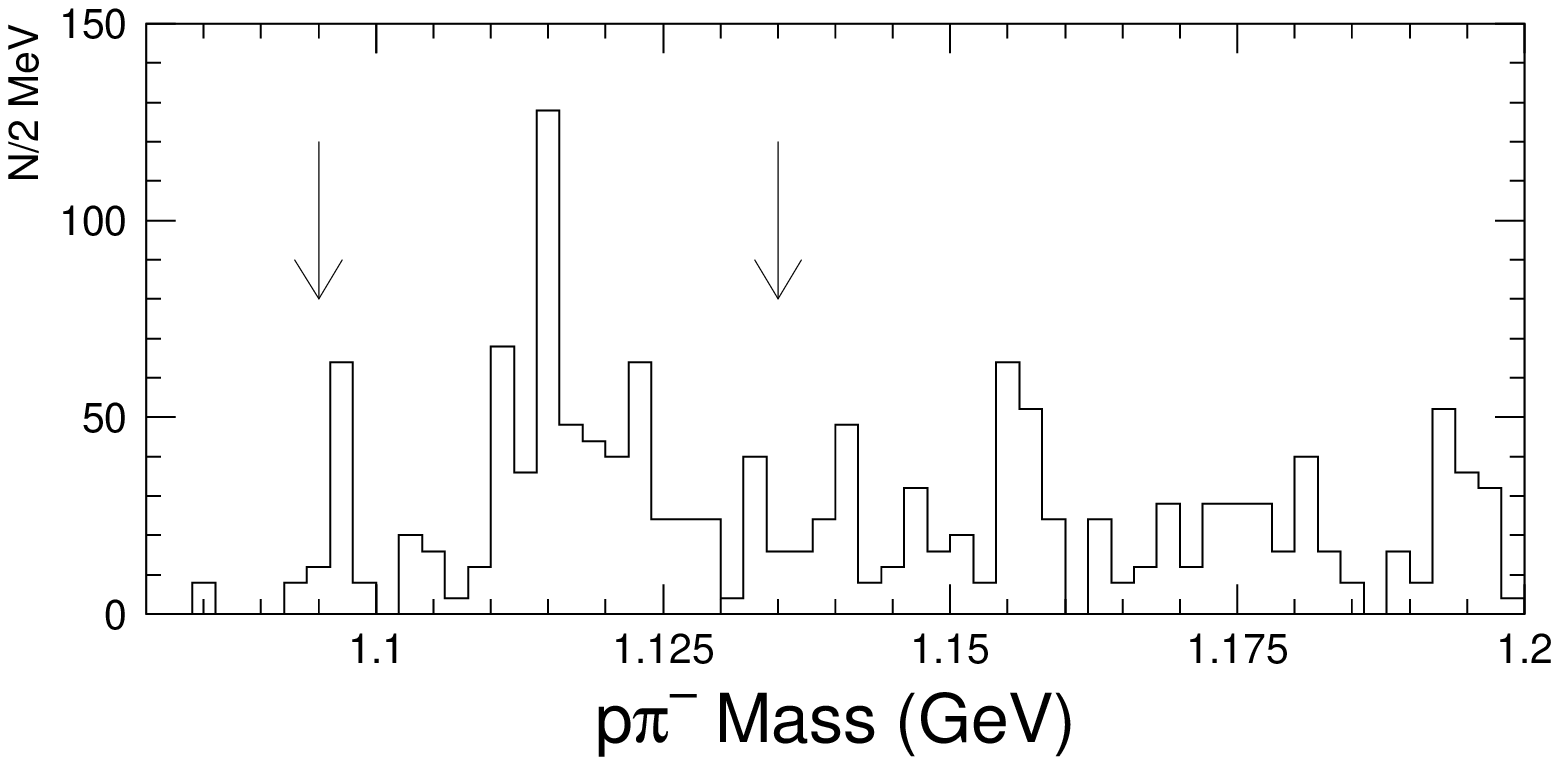}
  \caption{\label{fig:xis_lam}
         Distribution of $p\pi^-$ masses from a kinematic fit to
         $\psi(2S)\to p\overline{p}\pi^+\pi^-\pi^+\pi^-\pi^+\pi^-$
         showing the $\Lambda$ peak in
         $\Xi{}^0(1530)\overline{\Xi}{}^0(1530)$ candidate events.}
\end{figure}

As the dominant decay mode in $\Xi^*$ decays includes a $\Lambda$ in
the decay chain, a loose cut is placed on the $p\pi^-$ mass
($|M_{p\pi^-}-M_\Lambda|<20~\mbox{MeV}/c^2$) to enhance the signal fraction
(Figure~\ref{fig:xis_lam}).  Due to $\pi$ combinatorics, each
event that passes the kinematic fit is counted four times in this
plot. 

\begin{figure}
  \centering
  \includegraphics[width=0.90\linewidth]{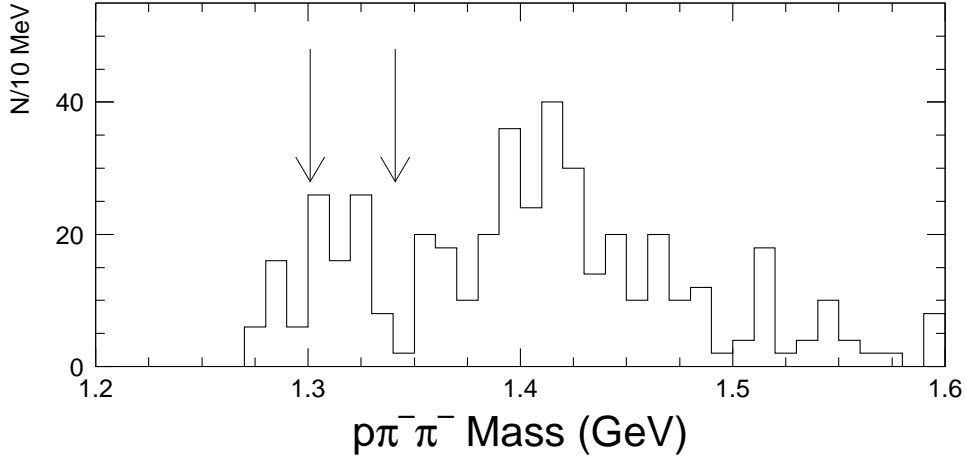}
  \caption{\label{fig:xis_xi}
         Distribution of $p\pi^-\pi^-$ masses from a kinematic fit to
         $\psi(2S)\to p\overline{p}\pi^+\pi^-\pi^+\pi^-\pi^+\pi^-$,
         showing the $\Xi^-$ peak in
         $\Xi(1530){}^0\overline{\Xi}(1530){}^0$ candidate events.}
\end{figure}

Similarly, as there is a $\Xi^-$ in the decay chain, a cut is made on
the $p\pi^-\pi^-$ invariant mass; $M_{p\pi^-\pi^-}$ is required to be
within 20~\mbox{MeV}$/c^2$ of the nominal mass of the $\Xi^-$, as
shown in Figure~\ref{fig:xis_xi}.  Due to $\pi$ combinatorics, each
event is counted twice in this plot.

\begin{figure}
  \centering
  \includegraphics[width=0.90\linewidth]{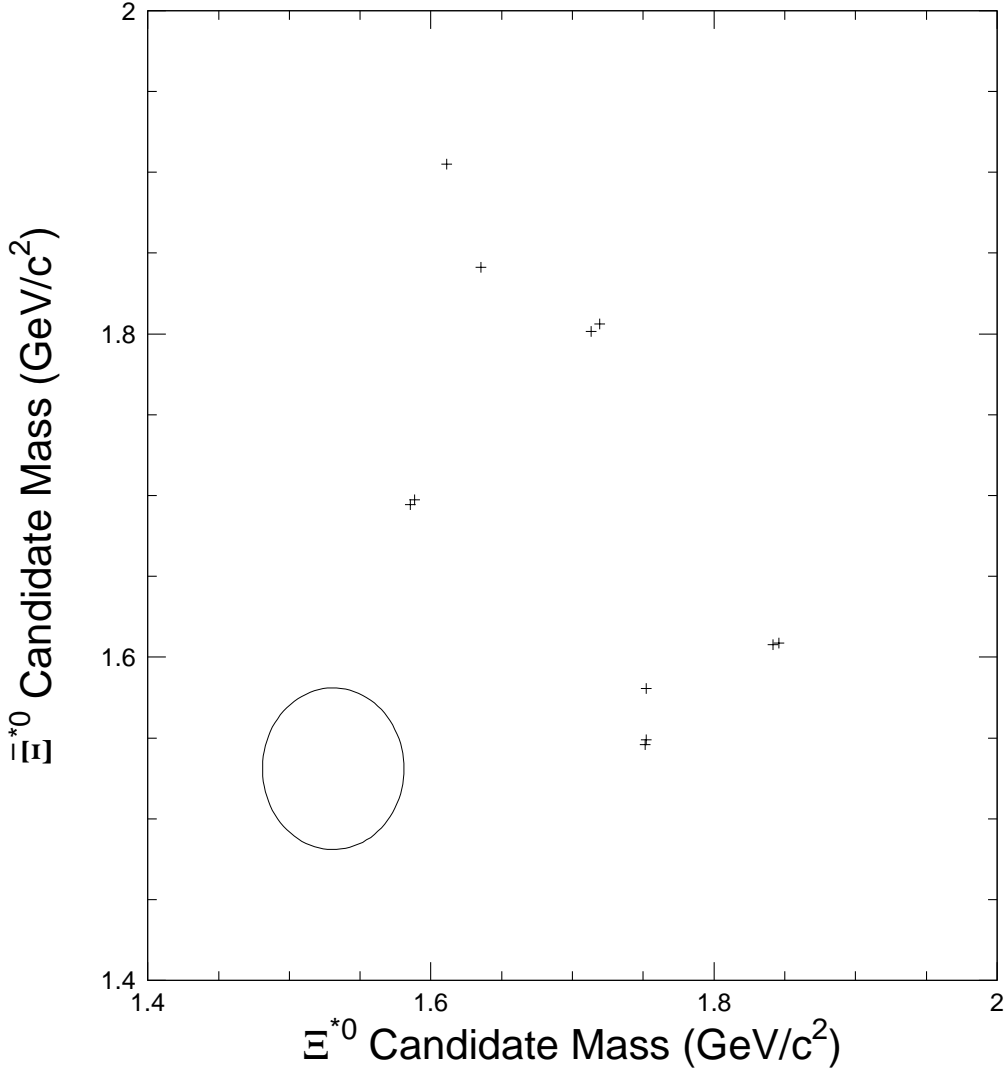}
  \caption{\label{fig:xis_mass}
         Distribution of $M_{\overline{p}\pi^+\pi^+\pi^-}$ vs 
         $M_{p\pi^-\pi^-\pi^+}$ from a kinematic fit to the final state
         $\psi(2S)\to p\overline{p}\pi^+\pi^-\pi^+\pi^-\pi^+\pi^-$.  
         The circle denotes the signal region, 3 sigma from the
         nominal mass of the $\Xi(1530)^0$.}
\end{figure}

All events which remain after the above cuts are graphed in 
Figure~\ref{fig:xis_mass} with $M_{\overline{p}\pi^+\pi^+\pi^-}$ on the 
vertical axis and $M_{p\pi^-\pi^-\pi^+}$ on the horizontal.  The signal 
region is shown as a circle at (1.531,1.531).  No events fall within the 
signal region defined as a 50~\mbox{MeV}$/c^2$ radius from the central 
value.  We set an upper limit of 2.3 events at 90\%\ CL for 
$N_{\mathstrut\Xi^*\overline{\Xi}{}^*}$.

\subsection{$\psi(2S) \to \Omega^-\overline{\Omega}{}^+$ }
The dominant $\Omega^-$ decay chain is
$\Omega^-\to\Lambda K^-$, $\Lambda\to p\pi^-$
with a total branching fraction of 43\%~\cite{PDG}.  We look
for $\psi(2S) \to \Omega^- \overline{\Omega}{}^+$
events with the topology $\psi(2S)\to p\overline{p}\pi^+\pi^-K^+K^-$,
i.e. six charged tracks where the 
$p\pi^-$ and $\overline{p}\pi^+$ are consistent with
being from the decay of a $\Lambda$ or $\overline{\Lambda}$.

We select events with six charged tracks in the polar angle region
$\left| \cos\theta \right| \leq 0.8$ and with zero net charge.  The
remaining events are subjected to a 7-constraint kinematic fit to the
hypothesis $\psi(2S) \to \Lambda \overline{\Lambda} K^+ K^-$,
$M_{\Lambda K^-} = M_{\overline{\Lambda} K^+}$.  The fit is applied for
each of the 36 particle assignment possibilities.  Only the assignment
with the best probability in the kinematic fit is considered.

\begin{figure}
  \centering
  \includegraphics[width=0.90\linewidth]{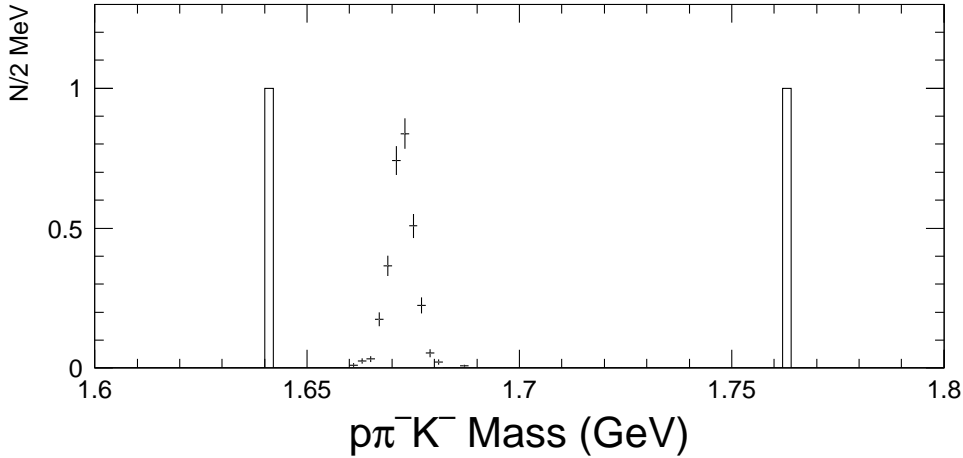}
  \caption{\label{fig:ommass} 
         Distribution of $\Lambda K^-$ masses from a kinematic fit to
         $\psi(2S)\to \Lambda\overline{\Lambda}K^+K^-$.  Histogram is
         candidate $\Omega^-\overline{\Omega}{}^+$ events, crosses are
         Monte Carlo.}
\end{figure}

Figure~\ref{fig:ommass} shows the $\Lambda K^-$ mass distribution for
the selected events, where the solid line histogram is data and the 
crosses are from Monte Carlo, normalized to three events. 
There are no candidates within three sigma of the nominal 
$\Omega^-$ peak, thus an upper limit of 2.3 is assigned at the 90
percent confidence level.

\section{Continuum background}

A few percent of the hadronic events in our data sample originate from
non-resonant $e^+e^-\to q\overline{q}$ annihilation events.  We use a
5.1~pb$^{-1}$ data sample taken off resonance to determine the level
of continuum $e^+e^-\to B_i\overline{B_i}$ contamination to our event
samples.  We find no events that survive the analysis procedures and
event selection criteria identical to those described above for either
of the modes $\Lambda\overline{\Lambda}$ or
$\Delta^{++}\overline{\Delta}{}^{--}$.  We conclude that continuum
events comprise a negligibly small contamination to our data samples.

\section{Determination of the total number of $\psi(2S)$ events}
\label{sec:npsi2S}

We determine the number of $\psi(2S)$ events in our data samples from
the observed number of cascade decays of the type
$\psi(2S)\to\pi^+\pi^- J/\psi$, $J/\psi\to X$.  The pions are
reconstructed, and the recoil mass of the two pions is fit to
determine the total number of $\psi(2S)\to\pi^+\pi^- J/\psi$ events.
From this fit, the total number of these events 
corrected for detection efficiency is $1.227 \pm 0.003 \pm 0.017
\times 10^6$.  The analysis for this is documented in
reference~\cite{leptonic}.

The total number of $\psi(2S)$ events is determined by dividing the
number of events in the previous paragraph by the PDG branching fraction
for the mode $\psi(2S)\to\pi^+\pi^- J/\psi$~\cite{PDG}.  This number
is determined to be $3.95 \pm 0.36 \times 10^6$ where the error is
dominated by the error on the $\psi(2S)\to\pi^+\pi^- J/\psi$ branching
fraction~\cite{leptonic_scale}.

\section{Acceptance and Efficiency}
\subsection{$p\overline{p}$}

We determine the efficiency for $\psi(2S)\to p\overline{p}$ events
from a sample of Monte Carlo simulated events.  Events were generated
with a distribution of $$\frac{dN}{d\cos\theta} \propto 1 +
\alpha\cos^2\theta$$ with $\alpha = 0.61\pm0.23$.  This proportionality
constant was measured in the $J/\psi \to p\overline{p}$ system by the
Mark II collaboration\cite{markii} and by the DM2
collaboration\cite{dm2}.  Out of 20000 events generated, 14857 events
survive these cuts, yielding a general efficiency of 0.743.  The
collinearity cut is also purely geometric, and has an efficiency of
0.999.

As the BES Monte Carlo is of limited usefulness for simulating
detailed hadronic interactions, cuts which are affected by such must
be corrected for by the examination of real data.  Fortunately, there
is a subset of events which allow the effects of these cuts to be
determined.  A clean sample of $p\overline{p}$ pairs was aquired from
the analysis of $\Delta^{++}\overline{\Delta}{}^{--}$ in
Section~\ref{sec:psip_dd}.  The $J/\psi$ contamination shown in
Figure~\ref{fig:dpp_psi} is the origin of this sample.  These events
are used to determine the $E_{SC}$, and $dE/dx$ cut efficiencies.

This study is summarized in Table~\ref{table:pp_eff}.  
The systematic error 
was determined from both Monte Carlo statistics and variation
of cuts.  The product of all efficiencies is: $0.234 \pm 0.022$.

\narrowtext
\begin{table}
\begin{center}
  \begin{tabular}{|c|r|r|r|r|r|}                                        \hline
  Cut           & $\epsilon_{\rm MC}$ 
                        & $\epsilon_{J/\psi}$
                                &  $\delta\epsilon$
                                        & $\delta{\cal B},{\rm eff}$
                                                & $\delta{\cal B},{\rm total}$
                                                                  \\ \hline
  General       & 0.743 &       & 0.006 & $0.3 \times 10^{-5}$  & $ 0.3 \times 10^{-5}$           \\ \hline
  Muon ID       & 0.696 &       & 0.007 & $0.2 \times 10^{-5}$  & $ 0.2 \times 10^{-5}$           \\ \hline
  BSC Geom      & 0.768 &       & 0.009 & $0.3 \times 10^{-5}$  & $ 0.3 \times 10^{-5}$           \\ \hline
  $E_{SC}<0.7$  &       & 0.610 & 0.048 & $1.8 \times 10^{-5}$  & $ 1.8 \times 10^{-5}$           \\ \hline
  $|XSP|<3$     &       & 0.968 & 0.047 & $1.1 \times 10^{-5}$  & $ 1.2 \times 10^{-5}$           \\ \hline
  Collinearity  & 0.999 &       & 0.001 & $0.02\times 10^{-5}$  & $ 0.02\times 10^{-5}$           \\ \hline
  \end{tabular}

  \caption{Relative efficiencies and systematic errors for cuts in the mode 
           $\psi(2S)\to p\overline{p}$ as modeled by Monte Carlo for
           geometric efficiencies and $J/\psi$ data for PID
           efficiencies.  Last column includes combined systematic
           error due to variation of cuts.}
  \label{table:pp_eff}
\end{center}
\end{table}
\widetext

\subsection{$\Lambda\overline{\Lambda}$, 
$\Sigma^0\overline{\Sigma}{}^0$, and $\Xi^-\overline{\Xi}{}^+$}

We determine the efficiency for the hyperon-pair channels completely
from Monte Carlo simulated events.  Here we generated 20000 events in
each mode with a $1+\alpha_d\cos\theta$ distribution, $\alpha_d = 0.67
\pm 0.21$, $0.22$, $0.5 \pm 0.5$ for $\Lambda\overline{\Lambda}$,
$\Sigma^0\overline{\Sigma}{}^0$, and $\Xi^-\overline{\Xi}{}^+$
respectively.  The value of $\alpha_d$ was not varied for the
$\Sigma^0\overline{\Sigma}{}^0$ mode as the statistical error was
large.  The values for $\alpha_d$ are those determined by the
Mark II collaboration\cite{markii} and by the DM2 collaboration\cite{dm2}.

The resulting efficiencies are summarized in Table~\ref{table:errors}.
The systematic error reported is a combination of Monte Carlo
statistics and variation of $\alpha_d$.  Also, an additional 10
percent error is added because of uncertainties in the kinematic
fitter used in these analyses.

In these three modes, we require two $\Lambda$'s that decay to charged
$p\pi$ final states, which have a branching fraction ${\cal B}
(\Lambda\to p\pi^-) = 0.639\pm0.005$; the other decay modes that are
required are very nearly unity, namely ${\cal
B}(\Sigma\to\Lambda\gamma) = 1.0$ and ${\cal B} (\Xi^-\to\Lambda\pi^-)
= 0.999$~\cite{PDG}.  The branching fraction acceptance for each
channel is the $\Lambda\to p\pi^-$ branching fraction squared:
$0.41\pm 0.01.$

\begin{table}
\begin{center}
\begin{tabular}{|l|c|c|c|c|} 
  mode    & $N_{evt}$           
                      & efficiency        & B.F. Acceptance  & KFit   \\ \hline
  $p\overline{p}$
          & $201 \pm 14 \pm 21$        
                      & $0.227 \pm 0.032$ & $1.00$           &        \\ \hline
  $\Lambda\overline{\Lambda}$
          & $80 \pm 9 \pm 4 $       
                      & $0.27\pm0.01$     & $0.41 \pm 0.01$  & 10\%   \\ \hline
  $\Sigma^0\overline{\Sigma}{}^0$
          & $8 \pm 3 \pm 2 $      
                      & $0.043\pm0.003$   & $0.41 \pm 0.01$  & 10\%   \\ \hline
  $\Xi^-\overline{\Xi}{}^+$
          & $12 \pm 3.4 \pm 0.2 $        
                      & $0.078\pm 0.01$   & $0.41 \pm 0.01$  & 10\%   \\ \hline
  $\Delta^{++}\overline{\Delta}{}^{--}$
          & $157 \pm 13 \pm 34$       
                      & $0.31\pm 0.02$    & $1.0 \pm 0.01 $  & 10\%   \\ \hline
  $\Sigma^{*+}\overline{\Sigma}{}^{*-}$
          & $14 \pm 4 \pm 3$          
                  & $0.104 \pm 0.005$ & $ 0.316 \pm 0.011 $  & 10\%   \\ \hline
  $\Xi^{*0}\overline{\Xi}{}^{*0}$
          & $< 2.3$             
                      & $0.041 \pm 0.001$ & $0.172 \pm 0.001$  & 10\% \\ \hline
  $\Omega^-\overline{\Omega}{}^+$
          & $< 2.3$             
                      & $0.042 \pm 0.001$ & $0.187 \pm 0.001$  & 10\% \\ \hline
\end{tabular}
  \caption{\label{table:errors} 
       Number of events, efficiencies, branching fraction acceptances,
       additional systematic errors due to the kinematic fit
       for $\psi(2S) \to B_i\overline{B_i}$.}
\end{center}
\end{table}

\subsection{Acceptance and Efficiency of the Decuplet Pairs}

We determine the efficiency for the decuplet hyperon-pair channels
completely from Monte Carlo simulated events.  Here we generated 20000
events in each mode with a $1+\alpha_d\cos\theta$ distribution,
$\alpha_d$ varying between 0 and 1 for
$\Delta^{++}\overline{\Delta}{}^{--}$, and constant at 0.6 for the
other modes.  The resulting efficiencies are summarized in
Table~\ref{table:errors}, where the systematic error reported is a
combination of Monte Carlo statistics and variation of $\alpha_d$.
Also, an additional 10 percent error is added because of uncertainties
in the kinematic fitter used in these analyses.

The branching fraction for $\Delta^{++}\to \pi^+\pi^+$ is greater than
0.99~\cite{PDG}, thus the branching fraction acceptance used is $1.00
\pm 0.01$.  The $\Sigma^{*+}\overline{\Sigma}{}^{*-}$ decay contains two
$\Lambda$s going to $p\pi$ ($0.639^2$) and two $\Sigma^{*}$s decaying
to $\Lambda\pi$ ($0.88^2$), for a total acceptance of $ 0.316 \pm
0.011$.  The $\Xi^{*0}\overline{\Xi}{}^{*0}$ decay has 3 components to the
acceptance: $\Xi^{*0}\to^-\pi^+$ ($0.650^2$), $\Xi^-\to\Lambda\pi^-$
($0.999^2$) $\Lambda\to p\pi^-$ ($0.639^2$), with a total acceptance
of $0.172 \pm 0.001$, and $\Omega^-\overline{\Omega}{}^+$ has only two
components in the acceptance, $\Omega^-\to\Lambda K^-$ ($0.678^2$) and
$\Lambda\to p\pi^-$ ($0.639^2$), with a total acceptance of $0.187 \pm
0.001$.

\section{Results}

The branching ratios 
${\cal B}(\psi(2S)\to B_i\overline{B_i})/{\cal B}(\psi(2S)\to J/\psi\pi^+\pi^-)$
are calculated by dividing the number of events in each mode, corrected
for efficiency and branching fraction acceptance, by the corrected number
of events in the reference mode, as noted in Section~\ref{sec:npsi2S}.
The final branching fractions are determined by multiplying the 
above branching ratios by the PDG value for 
${\cal B}(\psi(2S)\to J/\psi\pi^+\pi^-)$, $0.310 \pm 0.028$.
These are shown along with the branching ratios in Table~\ref{table:results}.

\begin{figure}
  \centering
  \includegraphics[width=0.90\linewidth]{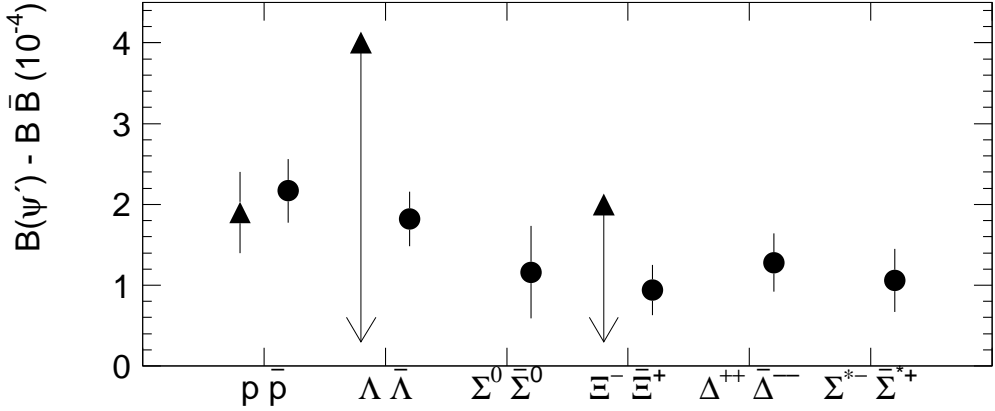}
  \caption{\label{fig:prev} 
         Comparison of measured branching fractions (circles) with
         previous measurements (triangles).  Two previous measurements
         are upper limits.}
\end{figure}

We compare our results for the branching fractions to previous limits
and results in Figure~\ref{fig:prev}.  Our measured value for the
${\cal B}(\psi(2S)\to p\overline{p})$ is about one standard deviation
higher than the previous DASP measurements, which was based on 4
events~\cite{dasp} and a Mark I measurement with similar
statistics~\cite{marki}.  The results for $\Lambda\overline{\Lambda}$
and $\Xi^-\overline{\Xi}{}^-$ are within the PDG upper limit values.
There are no previous experimental results for
$\Sigma^0\overline{\Sigma}{}^0$ or any of the decuplet modes.

\mediumtext
\begin{table}
\begin{center}
\begin{tabular}{|l|c|c|c|} 
  mode    & $N_{evt}$, Corr           
                      & ${\cal B}(B_i\overline{B_i})/
			  {\cal B}(J/\psi\pi^+\pi^-)~(\times 10^{-4})$
                                       & ${\cal B}(\times 10^{-5})$   \\ \hline
  $p\overline{p}$
          & $856 \pm 60 \pm 119$        
                      & $6.98 \pm .49 \pm .97 $
                                       & $21.6 \pm 1.5 \pm 3.6 $   \\ \hline
  $\Lambda\overline{\Lambda}$
          & $718 \pm 80 \pm 84 $       
                      & $5.85 \pm .65 \pm .69 $
                                       & $18.1 \pm 2.0 \pm 2.7 $   \\ \hline
  $\Sigma^0\overline{\Sigma}{}^0$
          & $456 \pm 162 \pm 152 $      
                      & $3.7  \pm 1.3 \pm 1.2 $
                                       & $12 \pm 4 \pm 4 $         \\ \hline
  $\Xi^-\overline{\Xi}{}^+$
          & $371 \pm 108 \pm 49 $        
                      & $3.0  \pm .9  \pm .4  $
                                       & $9.4  \pm 2.7 \pm 1.5 $   \\ \hline
  $\Delta^{++}\overline{\Delta}{}^{--}$
          & $506 \pm 40 \pm 127$       
                      & $4.12 \pm .33 \pm 1.04$
                                       & $12.8 \pm 1.0 \pm 3.4 $   \\ \hline
  $\Sigma^{*+}\overline{\Sigma}{}^{*-}$
          & $419 \pm 113 \pm 97$          
                      & $3.4  \pm .9  \pm .8  $
                                       & $11 \pm 3 \pm 3 $         \\ \hline
  $\Xi^{*0}\overline{\Xi}{}^{*0}$
          & $< 322$             
                      & $< 2.6                $
                                       & $< 8.1 $                  \\ \hline
  $\Omega^-\overline{\Omega}{}^+$
          & $< 290$             
                      & $< 2.4                $
                                       & $< 7.3 $                  \\ \hline
\end{tabular}
  \caption{\label{table:results}
       Numbers of events corrected for efficiency and branching
       fraction acceptance, branching fraction
       ${\cal B}(\psi(2S)\to B_i\overline{B_i})/$% 
       ${\cal B}(\psi(2S)\to J/\psi\pi^+\pi^-)$
       and final branching ratios for
       ${\cal B}(\psi(2S)\to B_i\overline{B_i})$.  Column 3 is
       calculated by dividing the corrected  number of events in each
       mode by the corrected number of events in the reference mode.}
\end{center}
\end{table}
\widetext

\begin{figure}
  \centering
  \includegraphics[width=0.90\linewidth]{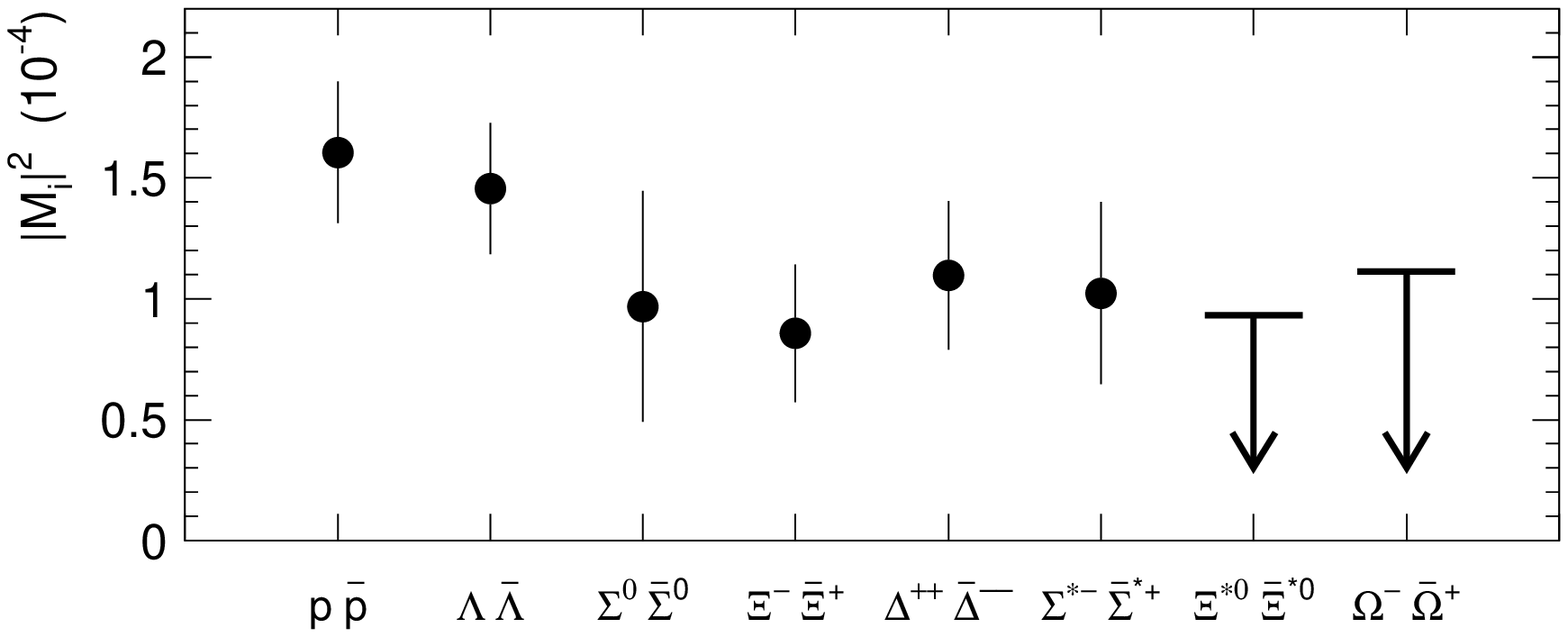}
  \caption{\label{fig:su3}
              The reduced branching fractions $\left|M_i\right|^2 =$
              ${\cal B}(\psi(2S)\to B_i\overline{B_i})
              /(\pi p^*/\sqrt{s})$ for
               $\psi(2S)\to B_i\overline{B_i}$ decays.}
\end{figure}

In Fig.~\ref{fig:su3}, we plot the reduced branching fractions derived
from our measurements.  The results show a trend to smaller values for
the higher massses, similar to that seen for the $J/\psi$ and are only
marginally consistent with expectations from flavor-$SU(3)$ symmetry.
Higher precision measurements {\it both} for the $J/\psi$ and
$\psi(2S)$ would clarify this issue.

\begin{figure}
  \centering
  \includegraphics[width=0.90\linewidth]{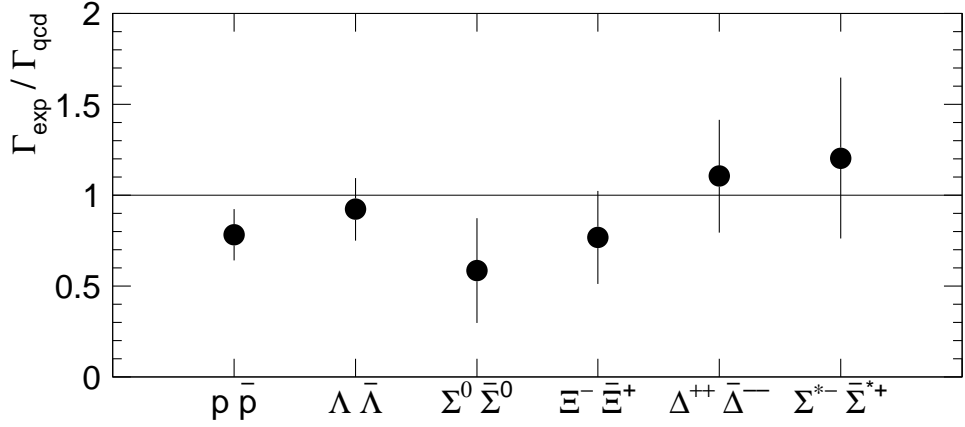}
  \caption{\label{fig:bolz}
         Comparison of ${\cal B}(\psi(2S)\to B_i\overline{B_i})$
         to Bolz and Kroll's predictions from perturbative 
         QCD.  Horizontal line is $\Gamma_{exp}/\Gamma_{qcd}=1.0$.}
\end{figure}

A comparison to the perturbative QCD predictions of Bolz and Kroll~\cite{bk}
is shown in Figure~\ref{fig:bolz}.  The results match quite well
with these calculations.

\begin{figure}
  \centering
  \includegraphics[width=0.90\linewidth]{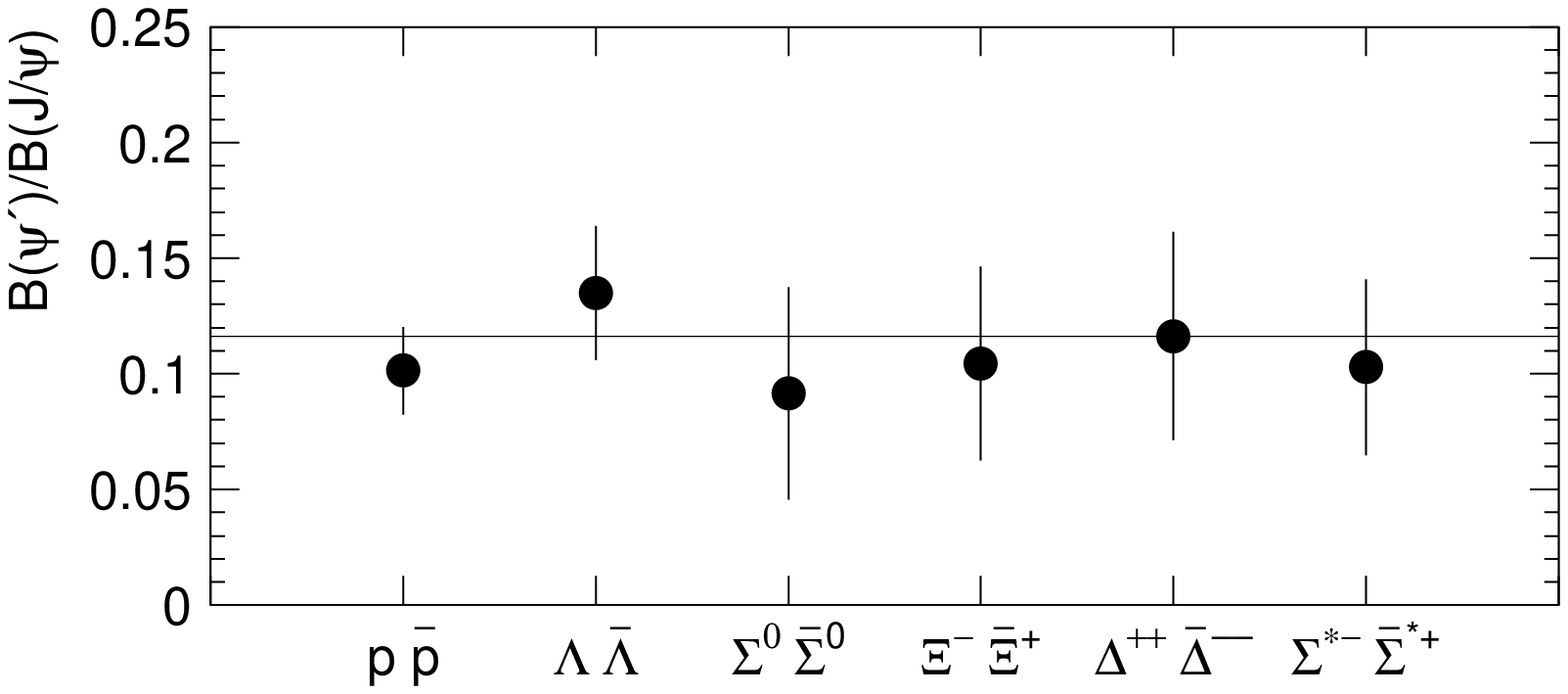}
  \caption{\label{fig:fourteen}
         The ratio ${\cal B}(\psi(2S)\to B_i\overline{B_i})/
           {\cal B}(J/\psi\to B_i\overline{B_i})$.  
         Horizontal Line is the 12 percent ratio expected from
         factorizing the $\psi(2S)\to B_i\overline{B_i}$ Feynman diagram.}
\end{figure}

Our measured $\psi(2S)$ branching fractions agree with expectations
derived from the application of the 12\% rule to the PDG values for
the corresponding $J/\psi$ decays for the modes $p\overline{p}$,
$\Lambda\overline{\Lambda}$, $\Sigma^0\overline{\Sigma}{}^0$
$\Xi^-\overline{\Xi}{}^-$, $\Delta^{++}\overline{\Delta}{}^{--}$ and
$\Sigma^{*-}\overline{\Sigma}{}^{*+}$, as shown in
Table~\ref{table:Psicomp} and in Figure~\ref{fig:fourteen}.  There are
no results for $J/\psi \to\Xi^0(1530)\overline{\Xi}{}^0(1530)$ and
$J/\psi\to\Omega^-\overline{\Omega}{}^+$ is not kinematically allowed.

\narrowtext
\begin{table}
  \begin{center}
  \begin{tabular}{|r@{$\to$}l|c|c|}
    \multicolumn{2}{|c|}{Decay Mode}  
         & \multicolumn{1}{c|}{$\cal B$}      
             & \multicolumn{1}{c|}{$0.116\times{\cal B}$ ($\times 10^{-5}$)}\\\hline
    $J/\psi$        & $p\overline{p}$ 
         & $(2.14\pm 0.10)\times 10^{-3}$
             & $ 24.8 \pm 1.2  $                                      \\ \hline
    $J/\psi$        & $\Lambda\overline{\Lambda}$
         & $(1.35 \pm 0.14)\times 10^{-3}$                   
             & $ 15.7 \pm 1.7 $                                       \\ \hline
    $J/\psi$        & $\Sigma^0\overline{\Sigma}{}^0$
         & $(1.3 \pm 0.2) \times 10^{-3} $ 
             & $ 15. \pm 2.  $                                        \\ \hline
    $J/\psi$        & $\Xi^-\overline{\Xi}{}^+$
         & $(0.9 \pm 0.2) \times 10^{-3} $ 
             & $ 10. \pm 2.  $                                        \\ \hline
    $J/\psi$        & $\Delta^{++}\overline{\Delta}{}^{--} $
         & $(1.10\pm 0.29)\times 10^{-3}$
             & $ 12.8 \pm 3. $                                        \\ \hline
    $J/\psi$        & $\Sigma^{*+}\overline{\Sigma}{}^{*-}$
         & $(1.03 \pm 0.13) \times 10^{-3} $ 
             & $ 11.9 \pm 01.5  $                                     \\ \hline
  \end{tabular}
  \caption{\label{table:Psicomp} 
           Branching ratio predictions for $\psi(2S)$.}
  \end{center}
\end{table}
\widetext

\section{Conclusions}
We report measurements of the branching fractions
for $\psi(2S)\to p\overline{p},~\Lambda\overline{\Lambda},~
\Sigma^0\overline{\Sigma}{}^0$, $\Xi^-\overline{\Xi}{}^+$,
$\Delta^{++}\overline{\Delta}{}^{--}$ and
$\Sigma{}^{+}(1385)\overline{\Sigma}{}^{-}(1385)$, along with upper limits
for the decays $\psi(2S)\to\Xi^0(1530)\overline{\Xi}{}^0(1530)$
and $\Omega^-\overline{\Omega}{}^+$.
The measured branching fractions agree with expectations 
based on an application of the 12\% rule to the corresponding
$J/\psi$ decays.  
The reduced branching fractions decrease with increasing baryon masses,
showing some deviation from expectations based on
flavor-$SU(3)$ symmetry.

\section{Acknowledgements}
The BES collaboration acknowledges financial support from the Chinese
Academy of Sciences, the National Natural Science Foundation of China, the
U.S. Department of Energy and the Ministry of Science \& Technology of Korea.
It thanks the staff of BEPC for their hard efforts.
This work is supported in part by the National Natural Science Foundation
of China under contracts Nos. 19991480 and 19825116
and the Chinese Academy of Sciences under contract No. KJ 95T-03(IHEP);
by the Department of Energy under Contract Nos.
DE-FG03-92ER40701 (Caltech), DE-FG03-93ER40788 (Colorado State University),
DE-AC03-76SF00515 (SLAC), DE-FG03-91ER40679 (UC Irvine),
DE-FG03-94ER40833 (U Hawaii), DE-FG03-95ER40925 (UT Dallas);
and by the Ministry of Science and Technology of Korea under Contract
KISTEP I-03-037(Korea).


\begin{thebibliography}{99}
\bibitem{hadratio}
$\alpha_s$ is calculated using the method given by the Particle Data
Group~\cite{PDG}, with $\Lambda^{(3)}=325~\mbox{MeV}$.  The ratio
${\alpha_s(\psi(2S))}/{\alpha_s(J/\psi)}$ is calculated to be about 0.8 by
this method.
\end{thebibliography}
\end{document}